\documentclass[aps,prl,showpacs,twocolumn,a4paper]{revtex4}
\usepackage{graphicx}
\usepackage{amsmath}
\usepackage{amsfonts}

\begin{document}

\title{Tuning Kinetic Magnetism of Strongly Correlated
Electrons via Staggered Flux}
\author{Yi-Fei Wang$^{1,2}$, Chang-De Gong$^{1,2}$, and Z. D. Wang$^{2,1}$}
\affiliation{$^1$National Laboratory of Solid State
Microstructures and Department of Physics,
Nanjing University, Nanjing 210093, China\\
$^2$Department of Physics and Center of Theoretical and
Computational Physics, The University of Hong Kong, Pokfulam Road,
Hong Kong, China}
\date{\today}

\begin{abstract}
We explore the kinetic magnetism of the infinite-$U$ repulsive
Hubbard models at low hole densities on various lattices with
nearest-neighbor hopping integrals modulated by a staggered
magnetic flux $\pm\phi$. Tuning $\phi$ from $0$ to $\pi$ makes the
ground state (GS) change from a Nagaoka-type ferromagnetic state
to a Haerter-Shastry-type antiferromagnetic state at a critical
$\phi_c$, with both states being of kinetic origin.
Intra-plaquette spin correlation, as well as the GS energy,
signals such a quantum criticality. This tunable kinetic magnetism
is generic, and appears in chains, ladders and two-dimensional
lattices with squares or triangles as elementary constituents.
\end{abstract}

 \maketitle

{\it Introduction.---}The study of magnetism may begin with a
Heisenberg-type model in which localized spins interact with each
other. However, the fundamental interaction between electrons is the
Coulomb-like repulsion, while the Pauli principle could induce a
spin dependence when electrons are not localized. The Nagaoka
theorem provides us a rigorous mechanism that the saturated
ferromagnetic (FM) state of kinetic origin is the unique ground
state (GS) when a single hole is inserted into the half-filled
Hubbard model with an infinite on-site repulsion $U$~\cite{Nagaoka}.
In the Nagaoka's problem, the sign of the hopping amplitudes around
the smallest closed loop in a lattice, ${\mathcal S}_{\text{loop}}$,
is necessarily positive.

In a recent letter, Haerter and Shastry (HS) have made important
progress in an opposite situation to Nagaoka's problem, a single
hole moving in the infinite-$U$ Hubbard model on two-dimensional
(2D) triangular lattices with frustrated hopping, i.e., ${\mathcal
S}_{\text{loop}}$ of an elementary triangle is
negative~\cite{Haerter}. Through numerical exact diagonalization
(ED) studies of finite lattices, they found that the motion of a
single hole with the electronic frustration leads to weak metallic
antiferromagnetism (AF) of kinetic origin.

There are still two important issues to be resolved. One is whether
there is a tunable transition between the two opposite kinetic
magnetisms. The other is whether or not the two kinetic magnetisms
can appear in the systems with more holes or with finite hole
densities, which are much closer to experimental realities.
Therefore, we are motivated to study the infinite-$U$ Hubbard models
at low hole densities on various lattices with nearest-neighbor
hopping integrals modulated by a staggered magnetic flux $\pm\phi$.
In these systems, ${\mathcal S}_{\text{loop}}=\exp({i\phi})$ can be
tuned from $+1$ to $-1$ via the Aharonov-Bohm effect, $\phi=0$ and
$\phi=\pi$ correspond respectively to Nagaoka's and HS's problems,
and the spatial periodicity is preserved when $\phi$ varies from $0$
to $\pi$.

\begin{figure}
\vspace{0.05in} \centering
\includegraphics[width=1.6in]{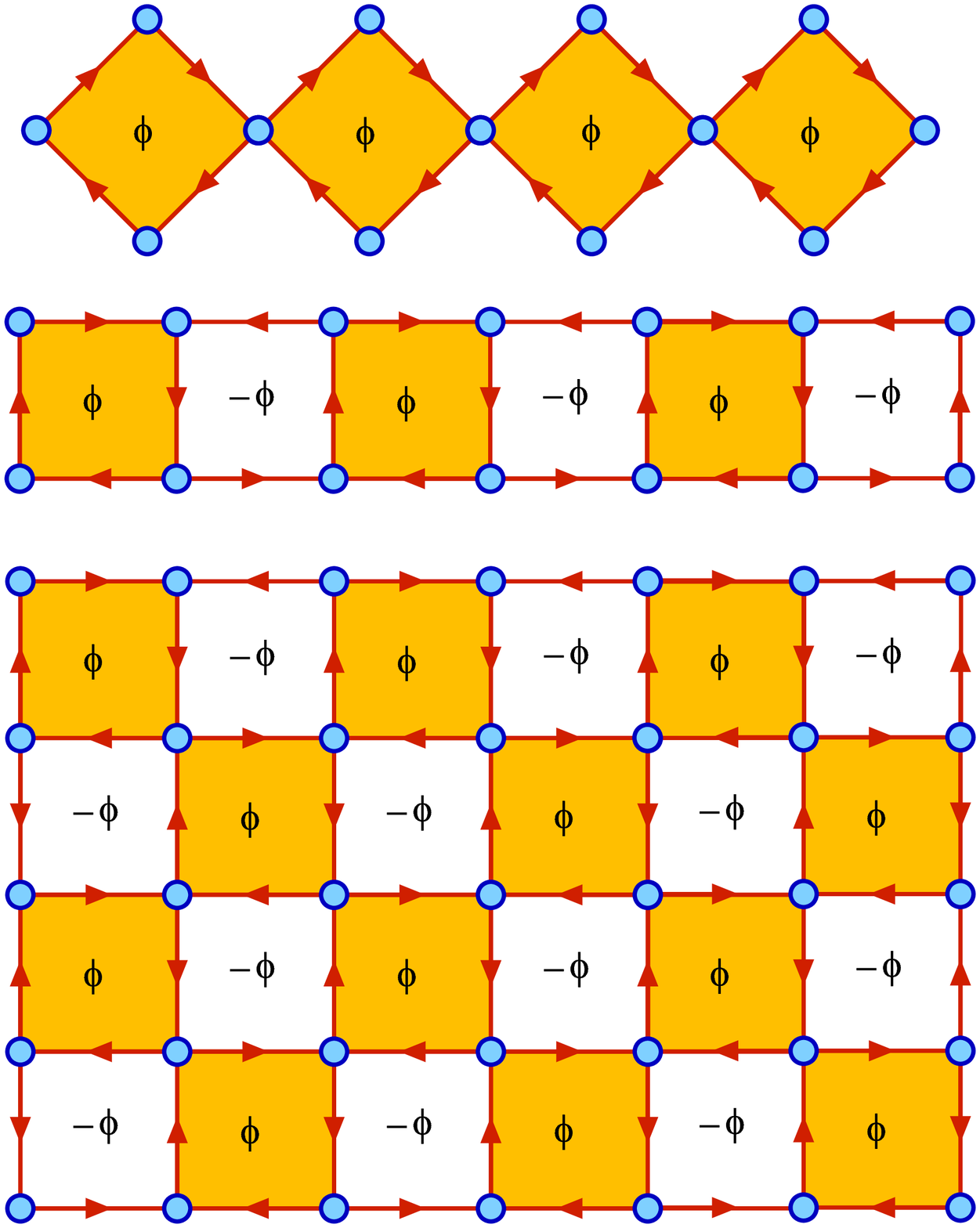}%
\hspace{-0.0in}%
\includegraphics[width=1.77in]{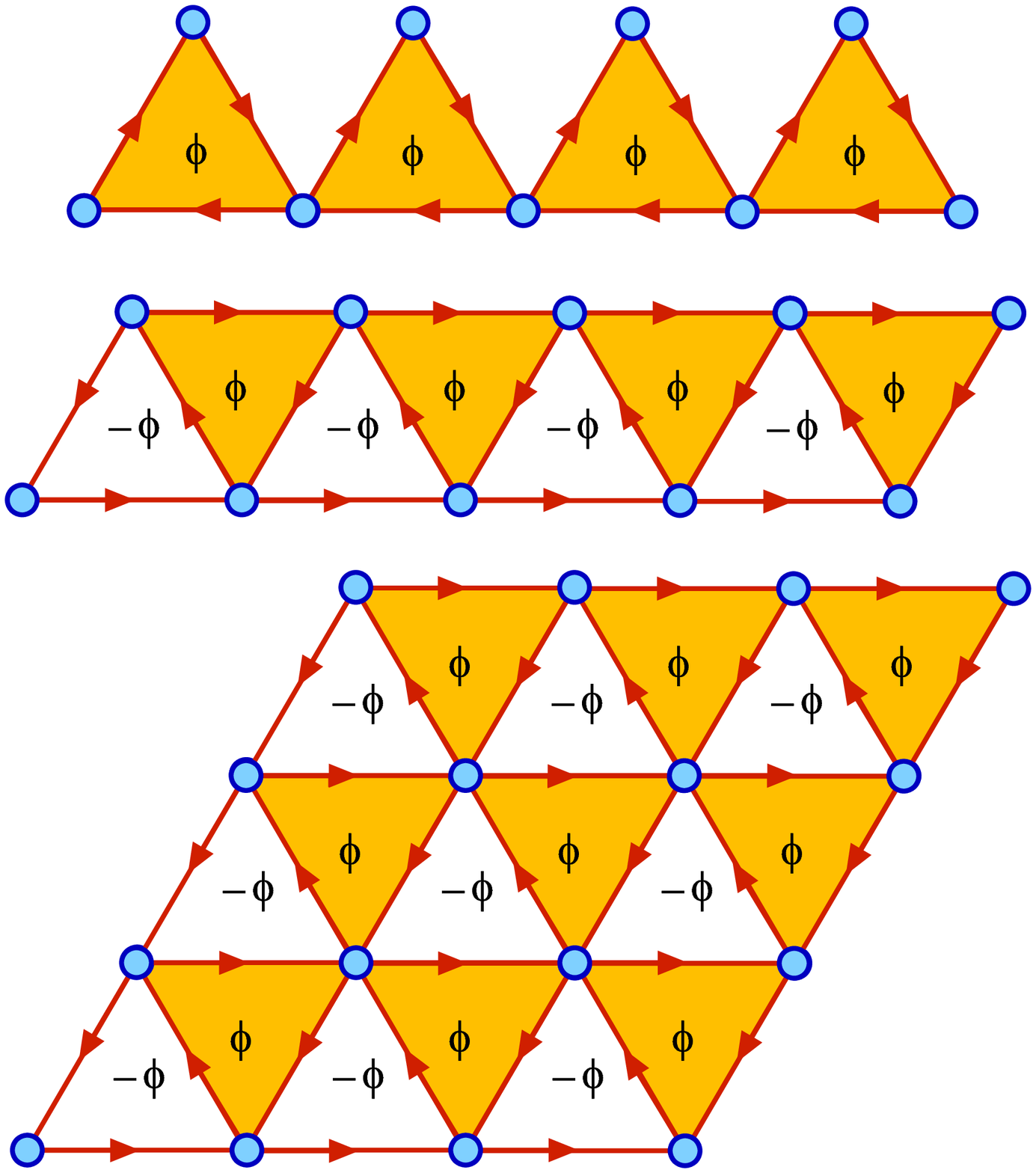}
\vspace{-0.1in} \caption{(color online). (left) Lattice structures
with squares as elementary plaquettes: diamond chain, square ladder
and 2D square lattice. (right) Lattice structures with triangles as
elementary plaquettes: sawtooth chain, trestle ladder and 2D
triangular lattice. A plaquette filled with color has a flux $\phi$,
and a blank one has a flux $-\phi$. Each arrow represents a phase
shift in its direction.}\label{f.1}
\end{figure}

Our motivations also come from three other aspects: our problem is
relevant to future experiments applying an external magnetic field
on artificial lattices of quantum dots~\cite{Kimura} or creating an
artificial magnetic field on optical lattices of ultracold fermionic
atoms~\cite{Demler}; an effective magnetic flux can be induced
intrinsically through some mechanisms in strongly correlated
systems~\cite{Taguchi,Affleck}; and our problem presents a concrete
example to study quantum phase transitions in both finite and
infinite systems~\cite{Sachdev} with a tunable parameter $\phi$.

On the basis of ED calculations of finite systems and analytical
estimations of infinite systems, we find that tuning $\phi$ from $0$
to $\pi$ makes the GS change from a Nagaoka-type FM state to an
HS-type AF state at a critical $\phi_c$. Intra-plaquette spin
correlation, as well as the GS energy, signals such a quantum
criticality. This tunable kinetic magnetism is generic, and appears
in many lattice structures with squares or triangles as elementary
constituents: elementary square and triangle, diamond and sawtooth
chains, square and trestle ladders, 2D square and triangular
lattices.

{\it Model Hamiltonian.---}The infinite-U Hubbard model with a
staggered flux (SF) can be written as:
\begin{equation}
H=t\sum_{\langle{ij}\rangle\sigma}e^{ia_{ij}}(1-n_{j,-\sigma})c^{\dagger}_{j\sigma}c_{i\sigma}(1-n_{i,-\sigma})+\text{H.c.}
\label{e.1}
\end{equation}
where the hopping integral $t$ is positive and is taken as the unit
of energy, $c_{i\sigma}$ ($c^{\dagger}_{i\sigma}$) is an electron
annihilation (creation) operator on site $i$ of spin
$\sigma=\uparrow$ or $\downarrow$, $n_{i\sigma}$ is the electron
number operator, and $\langle{ij}\rangle$ refers to two nearest
neighboring sites. The magnetic flux per plaquette (the summation of
$a_{ij}$ along four links around a plaquette) is given by $\pm\phi$
alternatively in neighboring plaquettes, with $\phi$ in units of
$\phi_0/2\pi$ ($\phi_0=hc/e$ is the flux quantum). Since the system
is symmetric under the transformations $\phi\rightarrow-\phi$ and
$\phi\rightarrow2\pi-\phi$, it is sufficient to restrict $\phi$ in
the interval $[0,\pi]$. The hole number is $N_h\equiv N_L-N_e$,
where $N_L$ and $N_e$ are the numbers of sites and electrons,
respectively; the hole density is denoted by $x\equiv N_h/N_L$.

The symmetric gauge is chosen for this SF as shown in Fig.
\ref{f.1}, and the corresponding periodical boundary conditions are
adopted. It should be noted that translational and macroscopic
time-reversal symmetries make our model distinct from the uniform
flux case, where the energy spectrum exhibits the fractal Hofstadter
butterfly and the uniform flux induces a saturated FM from
statistical transmutation at a flux quantum per
electron~\cite{Saiga}.

{\it Toy models of elementary square and triangle.---}As an
illustration of basic physics, we first consider toy models of
elementary square and triangle with a single hole ($N_h=1$) which
can be solved analytically. For the elementary square, when $\phi$
increases from $0$ to $\pi$, at $\phi_c=\pi/3$, the GS transits from
a state with the maximum total spin $S_{\text{tot}}={3\over2}$ (a
Nagaoka FM) to a state with $S_{\text{tot}}={1\over2}$, and the
nearest-neighbor (n.n.) spin correlation changes from ${1\over8}$ to
$-{1\over8}$. For the elementary triangle, when $\phi$ increases
from $0$ to $\pi$, at $\phi_c=\pi/2$, the GS transits from a
$S_{\text{tot}}=1$ state to a $S_{\text{tot}}=0$ state, and the n.n.
spin correlation changes from ${1\over{12}}$ to $-{1\over4}$.

\begin{table}
\caption{\label{t.1} Elementary square and triangle: GS properties
of the cases with $N_h=1$. }
\begin{tabular}{c|c|c|c}
\hline \hline \multicolumn{4}{c}{Elementary Square}\\
\hline
$\phi$ & $S_{\text{tot}}$ of GS & $\langle S_{i}\cdot S_{j}\rangle_{\text{n.n.}}$ & GS Energy \\
\hline
 $0\leq\phi<\pi/3$ & $3/2$ & $1/8$ & $-2\cos(\phi/4)$ \\
 $\pi/3<\phi\leq\pi$ & $1/2$ & $-1/8$ & $-2\cos(\phi/4-\pi/6)$ \\
\hline
\multicolumn{4}{c}{Elementary Triangle}\\
\hline
$\phi$ & $S_{\text{tot}}$ of GS & $\langle S_{i}\cdot S_{j}\rangle_{\text{n.n.}}$ & GS Energy \\
\hline
 $0\leq\phi<\pi/2$ & $1$ & $1/12$ & $-2\cos(\phi/3)$ \\
 $\pi/2<\phi\leq\pi$ & $0$ & $-1/4$ & $-2\cos(\phi/3-\pi/3)$ \\
\hline \hline
\end{tabular}
\end{table}

{\it Diamond and sawtooth chains.---}On a periodic lattice with an
SF, the lowest kinetic energy of a single hole in a saturated FM
spin background, $E^{\text{1h}}_{\text{FM}}(\phi)$, can be obtained
via the Fourier transformation. For an infinite diamond chain,
$E^{\text{1h}}_{\text{FM}}(\phi)=-2\sqrt{2}\cos(\phi/4)$; while for
a sawtooth chain,
$E^{\text{1h}}_{\text{FM}}(\phi)=-(1+\sqrt{5})\cos(\phi/3)$.

\begin{figure}[!htb]
  \vspace{-0.2in}
  \hspace{-0.27in}
  \includegraphics[scale=0.78]{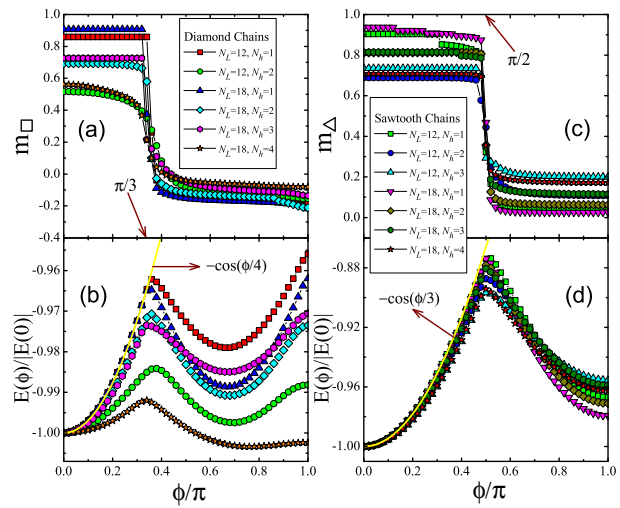}
  \vspace{-0.4in}
  \caption{(color online). (a),(b) Diamond chains. (c),(d) Sawtooth chains.
  $m$'s and rescaled GS energies versus $\phi$ in the cases with various
  sizes and hole numbers.} \label{f.2}
\end{figure}

We will present typical ED data obtained by using the Spinpack
package~\cite{Schulenburg} and exploiting the translational
invariance in the subspace with fixed $S^{z}_{\text{tot}}$ ($0$ for
even $N_e$, and ${1\over2}$ for odd $N_e$). The commonly used
quantity in evaluating Nagaoka FM, i.e. the GS $S_{\text{tot}}$, is
strongly dependent upon the boundary conditions chosen and the
even/odd parity of $N_{h}$~\cite{Saiga,Doucot,Troyer,Becca}. We
therefore concentrate on two derived quantities which are not
sensitive to the boundary condition or the parity of $N_{h}$.

The first quantity, $m$, is used to measure the spin correlations
intra a plaquette. For a square plaquette,
$m_{1}\equiv{1\over4}+\langle S_{2}\cdot S_{3}\rangle+\langle
S_{2}\cdot S_{4}\rangle+\langle S_{3}\cdot S_{4}\rangle$ (with four
clockwise sites $1$, $2$, $3$ and $4$, and $\langle\dots\rangle$
means the GS average) ~\cite{Joynt}, and the average over four sites
gives $m_{\square}={1\over4}\sum^{4}_{i=1}m_{i}$. On the square
lattice, $m_{\square}=1$ in a classical FM state, $m_{\square}=0$ in
a classical N\'{e}el AF state, while $m_{\square}={1\over4}$ if
there is no spin correlation. For a triangular plaquette,
$m_{1}\equiv{1\over2}+2\langle S_{2}\cdot S_{3}\rangle$, and the
average over three sites gives
$m_{\triangle}={1\over3}\sum^{3}_{i=1}m_{i}$. On the triangular
lattice, $m_{\triangle}=1$ in a classical FM state,
$m_{\triangle}={1\over4}$ in a classical 3-sublattice $120^{\circ}$
AF state, while $m_{\triangle}={1\over2}$ if there is no spin
correlation. The other quantity, the rescaled GS energy
$E(\phi)/|E(0)|$, is used to compare the non-analyticities in GS
energies of various cases.

For the $12$-site and $18$-site diamond chains with respectively
 $N_h=1-2$ and $N_h=1-4$, when $\phi$
changes from $0$ to $\pi$, $m_{\square}$'s drop almost abruptly near
a $\phi_c\approx\pi/3$ [Fig.~\ref{f.2}(a)]. (Note that
$\phi_c=\pi/3$ for the elementary square with $N_h=1$.) Meanwhile,
the GS energies also show clearly certain non-analyticities near
$\pi/3$ [Fig.~\ref{f.2}(b)].

For the $12$-site and $18$-site sawtooth chains with respectively
$N_h=1-3$ and $N_h=1-4$, as seen from Figs.~\ref{f.2}(c) and (d),
the abrupt drops of $m_{\triangle}$'s and non-analyticities in GS
energies occur near a $\phi_c\approx\pi/2$. (Note that
$\phi_c=\pi/2$ for the elementary triangle with $N_h=1$.)

In the four cases of diamond chains ($N_L=12$ with $N_h=1$, $N_L=18$
with $N_h=1-3$) and all seven cases of sawtooth chains, $m$'s change
very little (and the systems possess intra-plaquette FM correlations
since all $m$'s satisfy $m>{1\over2}$) when $\phi$ varies from $0$
to $\phi_c$, and the curves of rescaled GS energies are very close
to that of
$E^{\text{1h}}_{\text{FM}}(\phi)/|E^{\text{1h}}_{\text{FM}}(0)|$
[the continuous curves in Figs.~\ref{f.2}(b) and (d)]. (The ED data
also tell us that in each case with odd $N_{h}$, the GS
$S_{\text{tot}}$ takes the maximum value $N_{e}/2$ when
$0\leq\phi\ll\phi_c$.) Since $\phi=0$ corresponds to the Nagaoka FM
(${\mathcal S}_{\text{loop}}=+1$) and $m$'s change little for
$0\leq\phi<\phi_c$, we extend the notion of Nagaoka FM here and such
a state (with $0\leq\phi<\phi_c$) is referred to as a Nagaoka-type
FM. When $\phi$ varies from $\phi_c$ to $\pi$, all $m$'s show
obvious intra-plaquette AF correlations ($m_{\square}<{1\over4}$ for
diamond chains, and $m_{\triangle}<{1\over2}$ for sawtooth chains).
Since $\phi=\pi$ corresponds to the HS's problem (${\mathcal
S}_{\text{loop}}=-1$) and $m$'s change little as
$\phi_c<\phi\leq\pi$, we also extend the notion of HS AF and such a
state (with $\phi_c<\phi\leq\pi$) is referred to as an HS-type AF.

{\it Square and trestle ladders.---}For the single-hole Nagaoka FM
in an infinite square ladder with an SF,
$E^{\text{1h}}_{\text{FM}}(\phi)=-\sqrt{5+4\cos(\phi/2)}$; while in
an infinite trestle ladder,
$E^{\text{1h}}_{\text{FM}}(\phi)=-4\cos(\phi/3)$.

\begin{figure}[!htb]
  \vspace{-0.2in}
  \hspace{-0.27in}
  \includegraphics[scale=0.78]{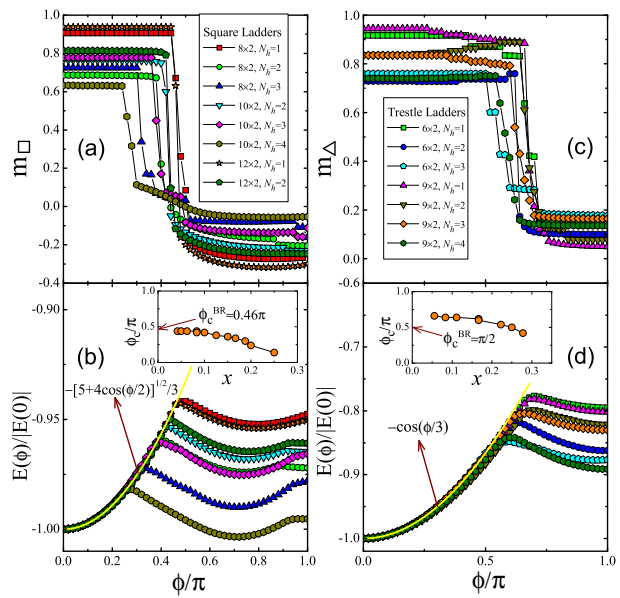}
  \vspace{-0.4in}
  \caption{(color online). (a),(b) Square ladders. (c),(d) Trestle ladders.
  $m$'s and rescaled GS energies versus $\phi$ in the cases with various
  sizes and hole numbers. The insets in
  (b) and (d) show $\phi_c$ vs. hole density $x$.} \label{f.3}
\end{figure}

For the $8\times2$, $10\times2$ and $12\times2$ square ladders
with respectively $N_h=1-3$, $N_h=2-4$ and $N_h=1-2$, as seen from
Figs.~\ref{f.3}(a) and (b), the abrupt drops of $m_{\square}$'s
and non-analyticities in GS energies occur near different
$\phi_c$'s, with $\phi_c$ versus hole density $x$ varying rather
smoothly.

For the $6\times2$ and $9\times2$ trestle ladders with
respectively $N_h=1-3$ and $N_h=1-4$ [Figs.~\ref{f.3}(c) and (d)],
the abrupt drops of $m_{\triangle}$'s and non-analyticities in GS
energies  also occur near different $\phi_c$'s, with still a
smooth curve of $\phi_c$ versus $x$.

In these cases of square and trestle ladders, $m$'s change very
little and satisfy $m>{1\over2}$ when $\phi$ varies from $0$ to
$\phi_c$, the curves of rescaled GS energy are very close to that of
$E^{\text{1h}}_{\text{FM}}(\phi)/|E^{\text{1h}}_{\text{FM}}(0)|$,
and these states are of Nagaoka-type FMs; when $\phi$ varies from
$\phi_c$ to $\pi$, all $m$'s show obvious intra-plaquette AF
correlations, and these states are of HS-type AFs.

It is tempting to estimate $\phi_c$ in the limit of low hole density
($x\rightarrow0$). Such a task can be partly fulfilled with the
Brinkman-Rice (BR) approximation~\cite{Brinkman}. For a single hole
in an infinite N\'{e}el AF spin background, the BR approximation
accounts the dominant contributions to the self energy of
single-hole Green's function, i.e., the retraceable paths without
any closed loop. Such an approximation leads to a hole band edge
(i.e., the lowest single-hole kinetic energy)
$E^{\text{BR}}_{\text{AF}}=-2\sqrt{z-1}$, where $z$ is the
coordination number.

Through comparison between $E^{\text{BR}}_{\text{AF}}$ and
$E^{\text{1h}}_{\text{FM}}(\phi)$ of a single hole in infinite
ladders, one can obtain a rough estimation of $\phi_c$ in the limit
$x\rightarrow0$. For an infinite square ladder, $z=3$, and
$E^{\text{BR}}_{\text{AF}}= E^{\text{1h}}_{\text{FM}}(\phi)$ gives a
$\phi^{\text{BR}}_c=2\arccos(3/4)\approx0.46\pi$. While for an
infinite trestle ladder, $z=4$, and $E^{\text{BR}}_{\text{AF}}=
E^{\text{1h}}_{\text{FM}}(\phi)$ gives a $\phi^{\text{BR}}_c=\pi/2$.

{\it Square and triangular lattices.---}For the single-hole Nagaoka
FM in an infinite square lattice with an SF,
$E^{\text{1h}}_{\text{FM}}(\phi)=-4\cos(\phi/4)$; while for an
infinite triangular lattice,
$E^{\text{1h}}_{\text{FM}}(\phi)=-6\cos(\phi/3)$.

\begin{figure}[!htb]
  \vspace{-0.2in}
  \hspace{-0.26in}
  \includegraphics[scale=0.78]{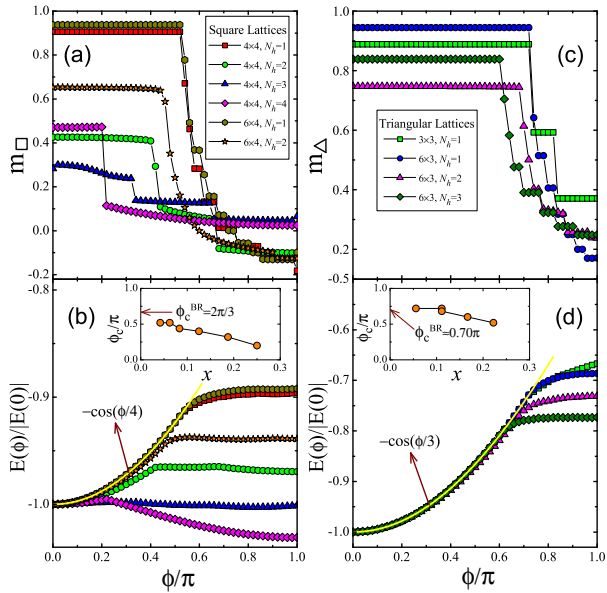}
  \vspace{-0.4in}
  \caption{(color online). (a),(b) Square lattices. (c),(d) Triangular lattices.
  $m$'s and rescaled GS energies versus $\phi$ in the cases with various
  sizes and hole numbers. The insets of
  (b) and (d) show $\phi_c$ vs. hole density $x$.} \label{f.4}
\end{figure}

For the $4\times4$ and $6\times4$ square lattices with respectively
$N_h=1-4$ and $N_h=1-2$  [Figs.~\ref{f.4}(a) and (b)], or the
$3\times3$ and $6\times3$ triangular lattices with respectively
$N_h=1$ and $N_h=1-3$  [Figs.~\ref{f.4}(c) and (d)], the abrupt
drops of $m$'s and non-analyticities (or changes of concavities) in
GS energies also occur near different $\phi_c$'s. In all these
cases, the GSs are of Nagaoka-type FMs for $0\leq\phi<\phi_c$, and
are of HS-type AFs for $\phi_c\ll\phi\leq\pi$. The curve of $\phi_c$
vs. $x$ of the square lattices [the inset in Fig.~\ref{f.4}(b)]
approaches a critical doping $x_c\sim0.3$ at $\phi_c=0$, which
agrees well with the quantum Monte Carlo studies on the instability
of saturated Nagaoka FM against doping~\cite{Becca}.

A rough estimation of $\phi_c$ in the limit $x\rightarrow0$ with the
aid of the BR approximation is: $\phi^{\text{BR}}_c=2\pi/3$ for an
infinite square lattice, and
$\phi^{\text{BR}}_c=3\arccos(\sqrt{5}/3)\simeq0.70\pi$ for an
infinite triangular lattice.

{\it Long-range spin correlations and ordering.---}As seen from the
above, the two quantities, $m$ and $E(\phi)/|E(0)|$, describe well
the transitions from the Nagaoka-type FM to the HS-type AF. The ED
data of the six kinds of lattices also tell us that for each case
with odd $N_{h}$, the GS $S_{\text{tot}}$ always takes the maximum
value $N_{e}/2$ and $\langle S_{i}\cdot S_{i+r}\rangle$ is positive
and almost a constant for any range $r$ when $0\leq\phi\ll\phi_c$,
namely, the Naogaka-type states are long-range ordered FM. Now we
would take a closer look at the long-range spin correlations of the
cases with even $N_h$'s. We focus on the three cases of square
ladders (in which there are the longest-range spin correlations) as
examples: $N_L=10\times2$ and $N_h=2,4$, $N_L=12\times2$ and
$N_h=2$.

\begin{figure}[!htb]
  \vspace{-0.2in}
  \hspace{-0.2in}
  \includegraphics[scale=0.6]{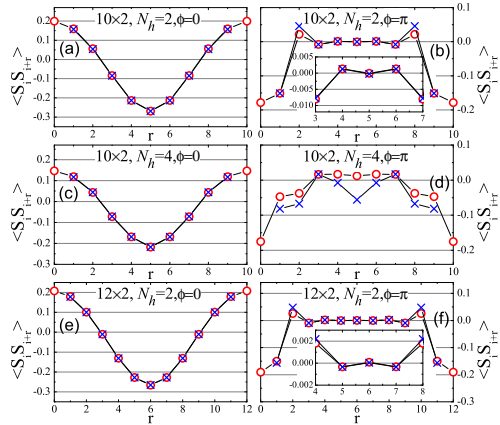}
  \vspace{-0.25in}
  \caption{(color online). Long-range intra-chain ($\times$) and inter-chain ($\circ$) GS spin correlations
  of square ladders at $\phi=0$ and $\phi=\pi$. The insets in (b) and (f) enlarge the middle parts.} \label{f.5}
\end{figure}

At $\phi=0$ [Figs.~\ref{f.5}(a),(c),and (e)], the GSs of the three
cases are singlets ($S_{\text{tot}}=0$), however, there is an
evidence of long-range FM in $\langle S_{i}\cdot S_{i+r}\rangle$ and
each singlet GS actually consists of two FM domains with opposite
magnetization as pointed out by Troyer {\it et al.}~\cite{Troyer}.
While at $\phi=\pi$ [Figs.~\ref{f.5}(b),(d),and (f)], the GSs have
decayed but extended (in the whole lattice) AF correlations.

{\it Summary and discussion.---}As for the infinite-$U$ repulsive
Hubbard models on the considered various lattices at low hole
densities ($x<0.3$) with n.n. hopping integrals modulated by a SF
$\pm\phi$, ED calculations of finite systems and analytical
estimations of infinite ones enable us to find: (i) tuning $\phi$
from $0$ to $\pi$ makes the GS change from a Nagaoka-type FM to an
HS-type AF at a critical $\phi_c$, and both states are of kinetic
origin; (ii) this tunable kinetic magnetism is generic, regardless
of the parity of $N_h$, and appears in various lattice structures
with squares or triangles as elementary constituents, i.e.
elementary plaquettes, chains, ladders and 2D lattices; (iii) near a
$\phi_c$, the intra-plaquette spin correlation drops abruptly from
FM to AF, and the GS energy exhibits non-analyticity; (iv) while the
$\phi_c$'s of chains are insensitive to $x$, the $\phi_c$'s of
ladders or lattices versus $x$ exhibit a decreasing and almost
smooth function; (v) the Nagaoka-type FM has long-range FM ordering,
while the HS-type AF has decayed but extended AF correlations.

In an artificial lattice of quantum dots, a large lattice constant
would enable us to observe this effect at a modest flux strength of
a few Tesla~\cite{Kimura}. This effect could also be realized in
optical lattices of ultracold atoms if appropriate phase factors are
introduced for hopping integrals by laser assisted hopping, lattice
tilting and other experimentally accessible
techniques~\cite{Demler}.

This work was supported by the NSFC and RGC grants of Hong Kong.

\end{document}